\documentclass[pra,aps,twocolumn]{revtex4}
\usepackage{amssymb}
\usepackage{latexsym}
\usepackage{graphicx}
\newcommand{\newsec}{\setcounter{equation}{0}\section}
\newcommand{\R}{{\mathbb R}}

\def\be{\begin{equation}}
\def\ee{\end{equation}}
\def\bea{\begin{eqnarray}}
\def\eea{\end{eqnarray}}

\def\d{{\,\rm d}}

\def\k{{\bf k}}

\def\q{{\bf q}}

\def\v{{\bf v}}

\def\bold0{{\bf 0}}

\begin{document}
\title{{\flushleft{\small {\rm Appeared in:
Phys. Rev. A {\bf 88}, 043640 (2013)}\\}}
\vspace{1cm} \large\bf Condensation of quasiparticles and density modulation beyond the superfluid critical velocity}
\author{Andr\'as S\"ut\H o$^1$ and P\'eter Sz\'epfalusy$^{1,2}$\\
$^1$Institute for Solid State Physics and Optics, Wigner Research Centre for Physics, Hungarian Academy of Sciences,\\ P. O. Box 49, H-1525 Budapest, Hungary\\
$^2$Department of Physics of Complex Systems, E\"otv\"os University, H-1117 Budapest, Hungary
}
\thispagestyle{empty}
\begin{abstract}
\noindent
We extend our earlier study of the ground state of a bosonic quasiparticle Hamiltonian by investigating the effect of a constant external velocity field. Below a critical velocity the ground state is a quasiparticle vacuum, corresponding to a pure superfluid phase at zero temperature. Beyond the critical velocity energy minimization leads to a macroscopic condensation of quasiparticles at a nonzero wave vector $\k_\v$ parallel to the velocity $\v$. Simultaneously, physical particles also undergo a condensation at $\k_\v$ and, to a smaller extent, at $-\k_\v$. Together with the Bose-Einstein condensation at $\k=\bold0$,  the three coexisting condensates give rise to density modulations of wave vectors $\k_\v$ and $2\k_\v$. For larger $|\v|$ our model predicts a bifurcation of $\k_\v$ with corresponding two pure condensates and no density modulation.

\vspace{2mm} \noindent PACS: 03.75.Nt, 05.30.Jp, 67.10.Ba, 67.25.dj

\vspace{2mm}
\noindent

\end{abstract}
\maketitle
%%%%%%%%%%%%%%%%%%%%%%%%%%%%%%%%%%%%%%  INTRODUCTION   %%%%%%%%%%%%%%%%%%%%%%%%%%%%%%%%%%%%%%%%%%%%%%%%%%

\newsec{Introduction}

Superfluid flow in systems of Bose-Einstein condensation (BEC) has been of great interest for a long time. Among the most interesting features is the existence of a critical velocity beyond which the motion is accompanied by dissipation even at zero temperature. The two main suggested mechanisms for the occurrence of a critical velocity have been the creation of quasiparticles (QPs) (Landau; cf. Ref.~\cite{LP}) and that of vortices (Feynman \cite{Fe}). Interestingly, it is possible to separate the contribution of QPs even if the true critical velocity due to vortex shedding is smaller than the Landau value \cite{WJCA}. The Landau criterion is relevant in case of channels of nanometer size, see e.g. \cite{ADPT}, and also for experiments where ions are dragged in liquid helium \cite{ACP}. The observed value of the critical velocity \cite{R...K}, \cite{O...K} has been attributed to vortex nucleation. It has been pointed out, however, that density inhomogeneity in a trapped Bose gas can also reduce considerably the Landau critical velocity as compared to that of the homogeneous gas \cite{FS}. A subsonic critical velocity has been derived in \cite{NG} using field theoretic methods. It is to be noted that a critical velocity in a trapped Fermi gas was also measured throughout the BEC-BCS crossover, and compared with the Landau criterion \cite{M...K}.

An interesting question is the structure of the fluid at velocities greater than the critical one. Within the Landau theory it was proposed that a roton condensate is created \cite{Pi}, \cite{Me}, which leads to a density modulation. The existence of density modulation was shown later also within the framework of Density Functional Theory \cite{ADPT}.

The purpose of the present work is to study QP condensation in a model, termed as the
Nozi\`eres-Saint James-Araki-Woods (NStJAW) scheme, and investigated earlier by us at $\v=\bold0$ flow velocity \cite{SSz}. At $\v=\bold0$ the ground state of the NStJAW quasiparticle Hamiltonian describes a superfluid at zero temperature. This is a QP vacuum state built up from a BEC of real particles in the $\k=\bold0$ one-particle state and from pairs of real particles in plane wave states of opposite nonzero momenta. In Sec. II we discuss the effect of an external velocity field on the ground state of this model. We find that for small velocities the ground state remains an unperturbed superfluid. When $|\v|$ exceeds the Landau critical value, the velocity field excites a macroscopic number of QPs of a single wave vector $\k_\v$ which is parallel to $\v$ and is determined by the energy minimum. The total momentum is carried by these QPs. The QP condensation at $\k_\v$ leads to the condensation of physical particles at $\k_\v$ and, to a smaller extent, at $-\k_\v$. The coexistence of the three condensates, those at $\k=\pm\k_\v$ and the original one at $\k=\bold0$, gives rise to a density modulation of wave vector $\k_\v$ and another one of $ 2\k_\v$. As the velocity increases, the condensate densities $n_\bold0$ and $n_{-\k_\v}$ decay to zero, and this may happen at different finite velocities. The $\k_\v$ and $ 2\k_\v$ density waves vanish with $n_\bold0$ and $n_{-\k_\v}$, respectively. Our model predicts a second critical velocity at which the solution for $\k_\v$ bifurcates. One of the solutions grows as $|\v|$, and the other one decays as $1/|\v|$ when $|\v|$ tends to infinity. A general ground state is a superposition of the corresponding two pure condensates. In Sec. III the density of the QP condensate is calculated in different approximations, for velocities close to the critical values. Section IV summarizes the results.

\newsec{Quasiparticle Condensation}

Recall our earlier definition \cite{SSz} of a quasi-particle Hamiltonian, %without the velocity field,
\bea\label{Hg}
H_{\rm QP}=w_0+\sum_\k e_\k M_\k+\sum_\k w_{\k\k}(M_\k^2-M_\k)\nonumber\\
+\sum_{\k\neq\k'}w_{\k\k'}M_\k M_{\k'}.
\eea
The summations run over $\k=\frac{2\pi}{L}(n_1,n_2,n_3)$, where $L$ is the side length of a cube of volume $V=L^3$ and $n_i$ are integers. $M_\k=b_\k^*b_\k$, and Bogoliubov's canonical transformation is applied in the form~\cite{VB}
\be
b_\k={1\over \sqrt{1-g_\k^2}}(a_\k-g_\k a_{-\k}^*)\qquad(\k\ne \bold0)
\ee
with $g_\k=g_{-\k}$ real, $-1<g_\k\leq 0$ (needed to minimize the vacuum energy); $a_\k,a_\k^*$ annihilate and create 'real' bosons, and $b_\k^*$ is the adjoint of $b_\k$.  For $\k=\bold0$ a shift replaces the Bogoliubov transformation: $b_0=a_0-z$ where $z$ is a real positive number of order $\sqrt{V}$ \cite{PH}.

The eigenstates of $H_{\rm QP}$ are eigenstates of the operators  $M_\k$ of the form
\be\label{basis}
\Phi_{(j_\k)}=\prod_\k\frac{1}{\sqrt{j_\k!}}(b_\k^*)^{j_\k}\Phi_0,
\ee
$(j_\k)$ being any terminating sequence of nonnegative integers. $\Phi_0$ is the common vacuum of all $b_\k$,
\be\label{dressed}
\Phi_0=e^{za_0^*}|0\rangle\otimes\left[\otimes_{\{\k,-\k\}}(1-g_\k^2)^{1/2}e^{g_\k a_\k^*a_{-\k}^*}|0\rangle\right],
\ee
where $|0\rangle$ is the physical vacuum and the second product runs over nonzero pairs.

In Ref.~\cite{SSz} we gave the expressions entering (\ref{Hg}). They depend on $g_\k$ via
\bea\label{hkchik}
h_\k&=&{g_\k^2\over 1-g_\k^2}=\langle \Phi_0|a_\k^*a_\k|\Phi_0\rangle,\nonumber\\
\chi_\k&=&{g_\k\over 1-g_\k^2}=\langle \Phi_0|a_\k a_{-\k}|\Phi_0\rangle,
\eea
so that $\chi_\k=-\sqrt{h_\k^2+h_\k}$. The energies $e_\k,w_{\k\k},w_{\k\k'}$ are positive, and we will not need the precise form of the vacuum energy $w_0$. In what follows, we approximate $v(\k)$, the Fourier transform of the pair potential, by $\nu=v(\bold0)$; the convergence of the infinite sums which appear in $w_0$ and $e_\k$ and involve $v(\k)$ will be ensured by the fast decay of $h_\k$ and $\chi_\k$. With this approximation, minimization of the vacuum energy with respect to $z$ and $\{g_\k\}$ results in
\be\label{ekNoz}
e_\k^2=\varepsilon(\k)^2+2\nu (n_0+n_a)\,\varepsilon(\k)+4\nu^2n_0n_a.
\ee
Here $\varepsilon(\k)=\hbar^2\k^2/2m$, and
\be\label{rho0'}
n_0=\frac{1}{V}\langle \Phi_0|a^*_0 a_0|\Phi_0\rangle=\frac{z^2}{V},\quad
n_a=\frac{1}{V}\sum_{\k\neq \bold0}|\chi_\k|.
\ee
We will also need
\be
n'=\frac{1}{V}\sum_{\k\neq \bold0}\langle \Phi_0|a_\k^*a_\k|\Phi_0\rangle=\frac{1}{V}\sum_{\k\neq \bold0}h_\k,
\ee
which is somewhat smaller than the density $n_a$ of the anomalous averages; however, $n_a$ goes to zero with $n'$ going to zero. From (\ref{ekNoz}), Bogoliubov's dispersion relation,
\be\label{ekBogo}
e_\k^2=\varepsilon(\k)^2+2\nu n_0\,\varepsilon(\k)
\ee
is obtained by assuming that $2\nu n_0n_a\ll\varepsilon(\k)(n_0+n_a)$ for the relevant values of $\k$, and $n_a\ll n_0$. We shall check the consistency of these assumptions. From Eqs.~(3.6) and (3.7) of Ref.~\cite{SSz},
\be\label{wkkBogo}
w_{\k\k}=\frac{\nu}{2V}\left(1+6h_\k+6h_\k^2\right)=\frac{\nu}{2V}\left(1+6\chi_\k^2\right)
\ee
and for $\k\neq \k'$,
\be\label{wkk'}
%w_{\k\k'}\approx 2 w_{\k\k}\approx 2w_{\k'\k'}.
w_{\k\k'}=\frac{\nu}{V}\left(1+2h_\k+2h_{\k'}+4h_\k h_{\k'}+2\chi_\k\chi_{\k'}\right).
\ee
Thus,
\be\label{wkk'ineq}
2\min\{w_{\k\k},w_{\k'\k'}\}\leq w_{\k\k'}\leq 2\max\{w_{\k\k},w_{\k'\k'}\}.
\ee
Note that $\varepsilon(\k)$ and $v(\k)$ depend only on $k=|\k|$. The same holds for $g_\k$ and, hence, for $h_\k$ and $\chi_\k$, if they are chosen so as to minimize the vacuum energy, and the minimizer is unique. Therefore, in Eqs.~(\ref{hkchik})-(\ref{wkk'}) we have functions of $k$ and modify the notations accordingly.

Now we introduce a constant velocity field in the quasi-particle Hamiltonian,
\bea
H_{\rm QP}(\v)=w_0+\sum_\k (e_k-\hbar\v\cdot\k) M_\k\nonumber\\
+\sum_\k w_{kk}(M_\k^2-M_\k)+\sum_{\k\neq\k'}w_{\k\k'}M_\k M_{\k'}
\eea
and look for the ground state of $H_{\rm QP}(\v)$. Let
\be
s_\k=\hbar \v\cdot \k-e_k+w_{kk}.
\ee
The eigenvalues of $H_{\rm QP}(\v)-w_0$ are
\be\label{eigenvalues}
%E_\v\{j_\k\}=\sum_\k j_\k[e_k-\hbar\v\cdot\k +w_{kk}(j_\k-1+2\sum_{\k'\neq\k}j_{\k'})].
E_\v\{j_\k\}=-\sum_\k s_\k j_\k+\sum_\k w_{kk}j_\k^2+\sum_{\k\neq \k'}w_{\k\k'}j_\k j_{\k'}.
\ee
Because $e_k$ starts with a constant [in (\ref{ekNoz})] or linearly [in (\ref{ekBogo})] at $k=0$, if $|\v|$ is small then $s_\k<0$ for each $k\neq 0$ (note: $w_{kk}\sim L^{-3}$); as a consequence, $\Phi_0$ remains the ground state ($j_\k\equiv 0$). Even if $|\v|$ is large, $s_\k$ is negative except for a finite number of $\k$: because $e_k$ grows quadratically with $k$, for any $\v\in\R^3$ the number of $\k$ vectors such that $s_\k>0$ is at most proportional to the volume. The eigenvalues with a single nonzero $j_\k$ have the form
\be
E_\v(\k,j_\k)=-s_\k j_\k+w_{kk}j_\k^2.
\ee
Supposing $s_\k>0$ and of order $L^0$, this can be negative, and its minimum is attained at $j_\k=m_\k$, where
\be\label{mk}
m_\k=\frac{s_\k}{2w_{kk}}\sim L^3
\ee
(more precisely, the closest integer to the value on the right). The corresponding eigenvalue is
\be
E_\v(\k,m_\k)=-\frac{s_\k^2}{4w_{kk}}<0;
\ee
it is also of the order of $V$, and still can be minimized with respect to $\k$. Because $m_\k$ is an integer and $\k$ also takes values on a lattice, the minimum may not be unique for all $\v$. To avoid this problem, we choose $\v$ parallel to a side of the cube, and exclude a discrete set of $v=|\v|$. Then, the unique minimum is attained at a $\k_\v$ parallel to $\v$:
\be\label{energy}
E_\v(\k_\v,m_{\k_\v})=-\frac{1}{4}\left[\max_k\frac{\hbar v k-e_k+w_{kk}}{\sqrt{w_{kk}}}\right]^2.
\ee
The corresponding eigenstate is
\be\label{GS}
\Phi_{m_{\k_\v}}=\frac{1}{\sqrt{m_{\k_\v}!}}(b_{\k_\v}^*)^{m_{\k_\v}}\Phi_0.
\ee
Below we show that this is actually the ground state of $H_{\rm QP}(\v)$. For a given $\v$ let ${\cal K'}$ denote the set of $\k$ vectors such that $s_\k>0$. We will suppose that
\be\label{bound}
\max_{\k\in{\cal K'}}w_{kk}<2\min_{\k\in{\cal K'}}w_{kk}\equiv 2w_{\cal K'}
\ee
%This assumption is equivalent with
%\be\label{bound2}
%\max_{\k\in{\cal K'}}\chi_k^2<\frac{1}{6}+2\min_{\k\in{\cal K'}}\chi_k^2\,;
%\ee
 which trivially holds in Bogoliubov's approximation. From (\ref{wkk'ineq}), for any $\k,\k'\in{\cal K'}$, $\k\neq\k'$,
\be\label{bound3}
w_{kk'}\geq 2w_{\cal K'}.
\ee
%The strategy of the proof is as follows.
In our search for the minimum of $E_\v\{j_\k\}$ we can set $j_\k=0$ for $\k$ outside ${\cal K'}$.
%We bound $E_\v\{j_\k\}$ from below by replacing $w_{kk'}$ with its minimum in ${\cal K'}$. We then minimize this lower bound with respect to each $j_\k$ for $\k\in{\cal K'}$, and prove that the minimum is just (\ref{energy}) if the unique element of ${\cal K'}$ is  $\k_\v$, and is strictly larger than (\ref{energy}) if ${\cal K'}$ contains at least two vectors.
We have
\be
E_\v\{j_\k\}\geq E_{\cal K'}\{j_\k\}
\ee
where
\be\label{lower}
E_{\cal K'}\{j_\k\}=-\sum_\k s_\k j_\k+\sum_\k w_{kk}j_\k^2+2w_{\cal K'}\sum_{\k\neq \k'}j_\k j_{\k'},
\ee
with summations over ${\cal K'}$. Define
\be
%J=\sum_\k j_\k,\quad
t_k=\frac{w_{\cal K'}}{2w_{\cal K'}-w_{kk}}.
\ee
With the assumption (\ref{bound}), $1\leq t_k<\infty$ for each $\k\in{\cal K'}$. Minimization of (\ref{lower}) w.r.t. each $j_\k$ yields
\be
j_\k=2t_k\sum_{\q\in{\cal K'}} j_\q -\frac{s_\k t_k}{2w_{\cal K'}}, \qquad\k\in{\cal K'}.
\ee
This set of equations has a unique solution for $j_\k$, and some of them may be negative. If this is the case, the set of $\k$ vectors must be restricted to a subset $\cal K''$ of $\cal K'$, $w_{\cal K'}$ replaced by $w_{\cal K''}$ and the minimization restarted. [Eqs.~(\ref{bound}-\ref{bound3}) are valid for ${\cal K''}\subset{\cal K'}$.] Let $\cal K$ be (any of) the largest subset(s) of $\cal K'$ such that for each $\k\in{\cal K}$ the solution of the minimization for $j_\k$ is positive. Let $m_\k$ be this solution. With the notation $M=\sum_{\cal K}m_\k$,
\be
m_\k=2t_k M -\frac{s_\k t_k}{2w_{\cal K}}
\ee
where $t_k$ is now defined with $w_{\cal K}$. From here, by summation over $\cal K$ we obtain
\be
M=\frac{1}{2w_{\cal K}(2\sum t_k-1)}\sum_\k s_\k t_k.
\ee
Insertion of the last two expressions into $E_{\cal K}\{m_\k\}$ results in
\be
E_{\cal K}\{m_\k\}=-\frac{\sum_{\k,\q}t_k t_q s_\k s_\q}{2w_{\cal K}(2\sum t_k-1)} + \frac{\sum_\k t_k s_\k^2}{4w_{\cal K}}.
\ee
Applying the inequality $s_\q s_\k\leq (s_\q^2+s_\k^2)/2$, we arrive at
\bea
E_{\cal K}\{m_\k\}&\geq& -\frac{\sum_\k t_k s_\k^2}{4w_{\cal K}(2\sum t_k-1)}\nonumber\\
&\geq& -\frac{\sum_\k t_k w_{kk}}{w_{\cal K}(2\sum t_k-1)}\max_{\k\in{\cal K}}\frac{s_\k^2}{4w_{kk}}
\nonumber\\
&\geq& \frac{\sum t_k w_{kk}}{w_{\cal K}(2\sum t_k-1)}E_\v(\k_\v,m_{\k_\v})\nonumber\\
&=&\left[1-\frac{\ell-1}{2\sum t_k-1}\right]E_\v(\k_\v,m_{\k_\v})
\eea
where $\ell$ is the number of vectors in $\cal K$. This shows that the minimum is attained with $\ell=1$. We thus conclude that the unique ground state of $H_{\rm QP}(\v)$ is (\ref{GS}) with eigenvalue $w_0+E_\v(\k_\v,m_{\k_\v})$. This conclusion is valid in a velocity interval whose lower edge is that $v$ beyond which $\max_k\{\hbar vk-e_k\}$ becomes positive; this value is identified with the superfluid critical velocity. We think that the condition (\ref{bound}) could be removed; however, the proof would be more lengthy.

It is useful to rewrite the above formulas in terms of densities. For a fixed $v$ greater than the critical velocity and any $k>e_k/\hbar v$, $\k$ parallel to $\v$, substituting $w_{kk}$ from Eq.~(\ref{wkkBogo}) we find
\be\label{sigmageneral}
\sigma_\k\equiv\lim_{V\to\infty}\frac{m_\k}{V}=\frac{\hbar v k-e_k}{\nu(1+6\chi_k^2)}
\ee
and
\bea\label{epsilonk}
\epsilon_v(k,\sigma_\k)&\equiv&
\lim_{V\to\infty}\frac{E_v(k,m_\k)}{V}\nonumber\\
&=&(e_k-\hbar v k)\sigma_\k+\frac{1}{2}\nu(1+6\chi_k^2)\sigma_\k^2\nonumber\\
&=&-\frac{(\hbar v k-e_k)^2}{2\nu(1+6\chi_k^2)}=-\frac{\nu}{2}(1+6\chi_k^2)\sigma_\k^2.\nonumber\\
\eea
Then, the equation corresponding to (\ref{energy}) is
\be\label{endens}
\epsilon_v(k_v,\sigma_{\k_\v})=-\frac{1}{2\nu}\left[\max_k\frac{\hbar vk-e_k}{\sqrt{1+6\chi_k^2}}\right]^2,
\ee
where $k_v=|\k_\v|$.

When varying $v$, the total density of physical particles must be kept fixed. Beyond the critical velocity this renders $\Phi_0$ and, thus, $n_0$, $n_a$ and $n'$ functions of $v$. The mean value of $N_\k=a_\k^*a_\k$ in $\Phi_{m_{\k}}$ can be obtained from
\be
N_\k=(1+h_k)M_\k+h_k M_{-\k}+h_k+\chi_k(b_\k b_{-\k}+b_\k^* b_{-\k}^*).
\ee
With the notation $\langle N_{\k'}\rangle_{m_\k}=\langle\Phi_{m_\k}|N_{\k'}|\Phi_{m_\k}\rangle$, for $\k\ne \bold0$ we have
\bea
\langle N_\k\rangle_{m_\k}&=&(1+h_k)m_\k+h_k,\nonumber\\
\langle N_{-\k}\rangle_{m_\k}&=&h_k m_\k+h_k,\nonumber\\
\langle N_{\k'}\rangle_{m_\k}&=&h_{k'}\quad (\k'\ne\pm\k).
\eea
Conservation of the density of physical particles implies
\be
n=n_0+n'+n_{\k_\v}+n_{-\k_\v}
\ee
where $n$ is the number density that we keep fixed,
\bea
n_{\k_\v}&=&\lim_{V\to\infty}\frac{1}{V}\langle N_{\k_\v}\rangle_{m_{\k_\v}}=(1+h_{k_v})\sigma_{\k_\v}\nonumber\\
n_{-\k_\v}&=&\lim_{V\to\infty}\frac{1}{V}\langle N_{-\k_\v}\rangle_{m_{\k_\v}}=h_{k_v}\sigma_{\k_\v}.
\eea
Thus, for (not too large) velocities beyond the critical value one has condensation of physical particles at $\k=\bold0$, $\k_\v$ and, to a smaller extent, at $-\k_\v$.

The coexistence of condensates with different wave vectors is accompanied by a density modulation. Indeed, in the Fourier transform of the density operator $\rho_\k=\sum_\q a_{\k+\q}^*a_\q$ we can replace, \emph{\`a la} Bogoliubov (and also rigorously~\cite{LSY}), $a_\bold0$ and $a_\bold0^*$ by $\sqrt{n_\bold0 V}$ and $a_{\pm\k_\v}$ and $a_{\pm\k_\v}^*$ by $\sqrt{n_{\pm\k_\v}V}$. Then, we obtain
\bea\label{densmod}
\frac{\|\rho_{\pm\k_\v}\Phi_{m_{\k_\v}}\|}{V}&\approx& \sqrt{n_\bold0 n_{\k_\v}}+\sqrt{n_\bold0 n_{-\k_\v}},\nonumber\\
\frac{\|\rho_{\pm 2\k_\v}\Phi_{m_{\k_\v}}\|}{V}&\approx& \sqrt{n_{\k_\v}n_{-\k_\v}}.
\eea
It is seen that the $\k_\v$ density modulation is due to the entanglement of the condensates at $\bold0$ and $\pm\k_\v$ and vanishes together with $n_\bold0$. On the other hand, the $ 2\k_\v$ density wave comes from the coexistence of the condensates at $\pm\k_\v$, and decays with $n_{-\k_\v}$.

Let us analyze the $v$-dependence of $\sigma_{\k_\v}$. This can be inferred from Eqs.~(\ref{sigmageneral})-(\ref{endens}) with the remark that the bounds
\be\label{sigmabound}
0\leq  \sigma_{\k_\v}=n_{\k_\v}-n_{-\k_\v}\leq n_{\k_\v}\leq n
\ee
must also be respected. It is easy to see that $\sigma_{\k_\v}$ attains $n$ at a finite velocity $v_1$.
Indeed, because $|\chi_k|$ tends to zero as $k$ increases, without the bound (\ref{sigmabound}) energy minimization would lead to the asymptotic (large $v$) results
$k_v=m v/\hbar$, $\sigma_{\k_\v}=m v^2/2\nu$ and $\epsilon_v(k_v,\sigma_{\k_\v})= -m^2v^4/8\nu$. With the bound (\ref{sigmabound}) we have, instead,
\be
\sigma_{\k_\v}\equiv n,\quad \epsilon_v(k_v,\sigma_{\k_\v})\equiv -\frac{\nu}{2}n^2\quad (v\geq v_1),
\ee
and the densities $n_0$, $n'$ and $n_{-\k_\v}$ vanish at respective velocities $v_0, v', v_-\leq v_1$ which may be different, and the largest of them equals $v_1$. If $v_0\neq v_-$,  the two density modulations (\ref{densmod}) disappear at different velocities. For $v\geq v_1$ the quasiparticles coincide with the physical ones, and $H_{\rm QP}(\v)$ goes over into the Hamiltonian of the so-called full diagonal model,
\bea
H_{\rm FD}(\v)&=&\sum_\k[\varepsilon(k)-\hbar\,\v\cdot\k] N_\k+\frac{\nu}{2V}(N^2-N)\nonumber\\
&+&\frac{1}{2V}\sum_{\k\neq \k'}v(\k-\k')N_\k N_{\k'},
\eea
studied earlier without the external velocity field \cite{FD}. Accordingly, $k_v$ is the solution for $k$ of the equation
\be\label{sigmasatured}
f(v,k)\equiv\hbar v k-\varepsilon(k)=\nu n\quad (v\geq v_1).
\ee
At $v=v_1$, $k_v$ still can be determined also from energy minimization. This provides a second equation,
\be\label{partialf}
\partial_k f(v,k)=0
\ee
which, together with Eq.~(\ref{sigmasatured}), can be used to compute $v_1$ and $k_1\equiv k_{v_1}$. Introducing
\be\label{crit}
c=\sqrt{\nu n/m},
\ee
the solution of Eqs.~(\ref{sigmasatured}) and (\ref{partialf}) is
\be
v_1=\sqrt{2}c,\quad \hbar k_1=m v_1.
\ee
In the actual model the saturation of the QP and energy densities occurs with a discontinuous derivative: the left-sided $v$-derivative of $\sigma_{\k_\v}$ and of $\epsilon_v(k_v,\sigma_{\k_\v})$ is nonzero at $v=v_1$, see also the end of Section 3.

The velocity $v_1$ not only marks density saturation, it is also a bifurcation point for $k_v$, see Figure 1. For $v>v_1$, $k_v$ is determined from Eq.~(\ref{sigmasatured}), and \emph{not} from energy minimization. The two solutions $k_\pm(v)$ are
\be\label{bifurc}
\frac{\hbar k_\pm}{m v}=1\pm\sqrt{1-2\left(\frac{c}{v}\right)^2}\quad (v>v_1).
\ee
In this way, at $v_1$ there is a (second) quantum phase transition: Between $v=0$ and $v=v_1$ the ground state of $H_{\rm QP}(\v)$ is unique, but there is a first quantum phase transition at the superfluid critical velocity ($c$ in Bogoliubov's approximation). For $v>v_1$, the ground states of $H_{\rm QP}(\v)$ form a two-dimensional subspace . In Eq.~(\ref{bifurc}) the plus sign corresponds to a pure condensate with an ever-increasing momentum density. The minus sign corresponds to a pure condensate with a vanishing momentum density $n\hbar k_-$, where $k_-\approx mc^2/v$ as $v$ tends to infinity. Along this solution momentum transfer to the system is a resonance phenomenon with a peak at $v_1$. The phase transition itself is subject to interpretation; it may signify the onset of turbulence. Note that there is no density modulation at velocities above $v_1$.

\begin{figure}
\centerline{\includegraphics[width=8cm]{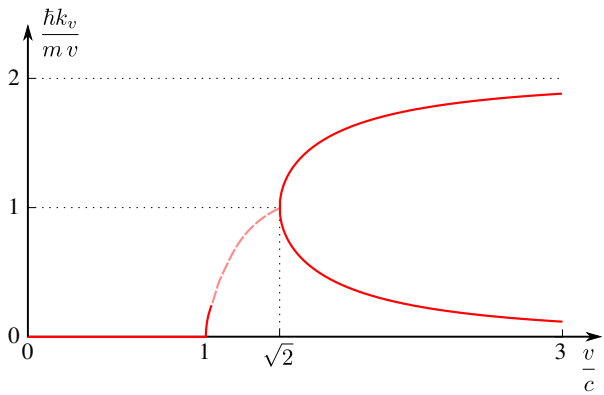}}
\caption{Variation of the wave number $k_v$ as a function of the velocity $v$, see Eqs.~(\ref{endens}) and (\ref{bifurc}).}
\end{figure}

\newsec{Quasiparticle condensate density close to the critical velocities}

The macroscopic condensation of quasiparticles at $\k_\v$ takes place independently of our use for $e_k$ of the gapful formula (\ref{ekNoz}) or of its Bogoliubov approximation (\ref{ekBogo}). We start the discussion using Eq.~(\ref{ekNoz}). In this case QP excitation is initiated by mode softening at a critical velocity $c'$ and wave number $k'>0$. While this occurs here due to the gap, an inflection point in the dispersion relation or a local minimum outside the origin (cf. roton mode) can also result in such a situation. At criticality $e_{k}=\hbar v k$ has a unique solution for $v$ and $k$:
\bea
c'&=&\sqrt{\frac{\nu(n_0+n_a)}{m}}\left(1+2\frac{\sqrt{n_0n_a}}{n_0+n_a}\right)^{1/2},\nonumber\\
k'&=&2\frac{\sqrt{\nu m}}{\hbar}(n_0n_a)^{1/4},
\eea
where $n_0$ and $n_a$ are the values at $v=0$, still unchanged at $v=c'$. If $v>c'$ then $k_v>k'$ can be large enough (for $n_0$ large) so that $h_{k_v}\approx 0$ hold true. Then $w_{k_v k_v}\approx \nu/2V$, and $k_v$ is obtained from the maximum of $\hbar v k-e_k$. To leading order in $v-c'$,
\be\label{kvgap}
k_v=k'+\hbar(v-c')/e_{k'}'',
\ee
\be\label{sigmagap}
\sigma_{\k_\v}=\frac{\hbar k'c'}{2\nu}\left(\frac{v}{c'}-1\right)
\ee
and $E_\v(\k_\v,m_{\k_\v})/V=-(\nu/2)\sigma_{\k_\v}^2$. The above treatment is meaningful if the interval $(c',v_1)$ is nonempty, that is, if $2\sqrt{n_0n_a}<n_0+2n'-n_a$, which holds if, say, $n_a/n_0<0.1$.

Next, we use the Bogoliubov approximation (\ref{ekBogo}) of $e_k$. This is based on the assumption that $n_0\approx n$ below the critical velocity. Since $e_k$ is a convex function of $k$ and
\be
e_k=\hbar k\sqrt{\nu n/m}+O(k^2)
\ee
near $k=0$, the critical velocity at which quasiparticle excitations appear is $c$ given by (\ref{crit}), see Fig.~1.

To minimize the energy density we need $\chi_k^2$. From Eqs.~(3.29), (4.14) and (4.20) of Ref.~\cite{SSz}, and assuming $n'\ll n_0$,
\be\label{hkneark=0}
h_k\approx \frac{\nu n_0+\varepsilon(k)}{2e_k}-\frac{1}{2}.
\ee
Here and below $n_0$ is the $v$-dependent value for $v>c$. Note that the convergence of $h_k$ to zero with either $k$ going to infinity or $\nu n_0$ going to zero can be seen on this formula. If $c<v\ll 2c$, $k_v$ will be close to zero and $n_0$ close to $n$, so
\be\label{hk}
h_k\approx-\chi_k\approx \frac{mc}{2\hbar k}.
\ee
Substituting (\ref{ekBogo}) and (\ref{hk}) into Eq.~(\ref{epsilonk}), keeping the terms of the order of $k^4$ and $k^6$ and minimizing with respect to $k$ we find
\be\label{k_v}
k_v=\sqrt{\frac{8}{3}}\frac{mc}{\hbar}\sqrt{\frac{v}{c}-1},
\ee
\be
\epsilon_v(k_v,\sigma_{\k_\v})=-\frac{64}{27}\,\nu n^2\left(\frac{v}{c}-1\right)^4,
\ee
and
\be\label{sig5/2}
\sigma_{\k_\v}=\frac{2}{3}\left(\frac{8}{3}\right)^{3/2}n\left(\frac{v}{c}-1\right)^{5/2}.
\ee
We still have to verify the consistency of the assumptions $2\nu n_0n_a\ll \varepsilon(k_v)(n_0+n_a)$ and $n_a\ll n_0$ which were at the origin of the Bogoliubov approximation. From (\ref{k_v}),
\be\label{varepsBogo1}
\varepsilon(k_v)=\frac{4}{3}mc^2\left(\frac{v}{c}-1\right)=\frac{4}{3}\nu n\left(\frac{v}{c}-1\right),
\ee
so the lower bound on the velocity reads
\be
\frac{v}{c}-1\gg \frac{3}{2}\frac{n_0n_a}{n(n_0+n_a)},
\ee
which is consistent with $v\ll 2c$ if $n_a\ll n_0$. This, however, holds true if the interaction is weak enough. There is also an absolute upper bound, $v/c-1\leq  3/2^{11/5}$, coming from $\sigma_{\k_\v}\leq n$ which can be read off from Eq.~(\ref{sig5/2}). However, the applicability of this formula does not extend up to this velocity.

For the Bogoliubov approximation at a somewhat larger $v$ but still close to $c$ we can again suppose $n_0\approx n$, $h_k=0$, $w_{k k}=\nu/2V$ to compute $k_v$ and $\sigma_{\k_\v}$. Thus,
\be\label{sigmavlarge}
\sigma_\k=\frac{\hbar v k-e_k}{\nu},
\ee
\be\label{epsilonvlarge}
\epsilon_v(k,\sigma_\k)=-(\nu/2)\sigma_\k^2
\ee
and $k_v$ is determined by the maximum of $\hbar v k-e_k$. This latter is attained for $\varepsilon=\varepsilon(k)$ satisfying the equation
\be\label{epsequa}
\varepsilon^2+(2mc^2-mv^2/2)\varepsilon=m^2c^2(v^2-c^2).
\ee
Suppose that
\be\label{epsiloncond}
\varepsilon\ll 2mc^2-mv^2/2.
\ee
Then
\be\label{epsilonapprox}
\varepsilon(k_v)=\frac{\hbar^2 k_v^2}{2m}\approx mc^2\frac{(v/c)^2-1}{2-(v/c)^2/2}
\ee
from which we get
\be\label{kv2}
k_v=\frac{\sqrt{2}mc}{\hbar\sqrt{1-(v/2c)^2}}\sqrt{\frac{v}{c}-1}.
\ee
Under the condition (\ref{epsiloncond}), $e_{k_v}\approx \hbar c k_v$, yielding
\be\label{sigmakv2}
\sigma_{\k_\v}\approx 2\sqrt{\frac{2}{3}}\frac{mc^2}{\nu}\left(\frac{v}{c}-1\right)^{3/2}
=2\sqrt{\frac{2}{3}}\,n\left(\frac{v}{c}-1\right)^{3/2}.
\ee
The interval of $v$ in which this law makes sense can be deduced from (\ref{epsiloncond}) and
(\ref{epsilonapprox}). From the first we see that $v<2c$; the comparison of the two gives $v\ll \sqrt{2} c=v_1$. Moreover, if $\alpha$ is the largest value of $v/c$ for the applicability of
(\ref{hkneark=0}) to $k=k_v$ given by (\ref{k_v}), and $\beta$ is the lowest value of $v/c$ for which $h_{k_v}=0$ is a good approximation \emph{and} (\ref{epsiloncond}) and (\ref{epsilonapprox}) are compatible, then both
\be\label{Bogo2}
1<\alpha<\beta<\sqrt{2},\qquad \frac{4}{3}\left(\alpha-1\right)<\frac{\beta^2-1}{2-\beta^2/2}
\ee
must hold. The second inequality follows from the monotonic growth of $k_v$ and, hence, of $\varepsilon(k_v)$ with $v$. At $\beta=\sqrt{2}$ the second inequality holds for $\alpha<1.75$, indicating that (\ref{Bogo2}) can easily be satisfied.

Somewhat more can be said about $\sigma_{\k_\v}$ if we suppose $h_k=\chi_k=0$ for all $v\geq c$. In general, between $c$ and $v_1$, $k_v$ is an invertible function of $v$. Let $v_k$ be its inverse. If we set $\chi_k=0$, then both $k_v$ and $v_k$ can be computed from the equation
\be
\partial_k \left[\hbar vk-\sqrt{\varepsilon(k)^2+2\nu n_0\varepsilon(k)}\right]=0.
\ee
Solving this equation for $v$, we find
\be\label{vk}
v_k=\sqrt{\frac{2}{m}}\frac{\varepsilon(k)+\nu n_0}{\sqrt{\varepsilon(k)+2\nu n_0}}.
\ee
Let
\be\label{sigmak}
\sigma(k)=\frac{1}{\nu}\left[\hbar kv_k-\sqrt{\varepsilon(k)^2+2\nu n_0\varepsilon(k)}\right];
\ee
this is just $\sigma_{\k_\v}$ if $k_v=k$. Substituting (\ref{vk}) into (\ref{sigmak}) and setting $n_0=n-\sigma(k)$ which is now the case, we arrive at the implicit equation
\be
\nu\sigma(k)=\frac{\varepsilon(k)^{3/2}}{[\varepsilon(k)+2\nu n-2\nu\sigma(k)]^{1/2}}.
\ee
From here, for $k$ small, i.e., $v$ close to $c$,
\be
\sigma_{\k_\v}=\frac{1}{\sqrt{2n}}\left(\frac{\varepsilon(k_v)}{\nu}\right)^{3/2},
\ee
to be compared with (\ref{kv2}) and (\ref{sigmakv2}). On the other hand, observing that $\nu n=\varepsilon(k_1)$, it is seen that for $v$ smaller than but close to $v_1$, $\sigma_{\k_\v}$ satisfies the equation
\be
\sigma_{\k_\v}=n-\frac{\hbar v_1}{\nu}(k_1-k_v)\approx n-\frac{m v_1}{\nu}(v_1-v),
\ee
so that
\be
\left(\frac{\d\sigma_{\k_\v}}{\d v}\right)_{v=v_1-0}=\frac{m v_1}{\nu}.
\ee

\section{Summary}

In this paper we applied a variational quasiparticle theory to study the ground state of a Bose system exposed to a constant external velocity field. We have shown that at small velocities the energy minimum for the variational ansatz occurs at zero quasiparticle excitation, meaning the persistence of a pure superfluid state at zero temperature. Crossing the Landau critical velocity a quasiparticle condensate is formed with spectacular consequences. The condensation takes place into a one-particle state of momentum $\k_\v$ which is parallel to the velocity $\v$ and whose magnitude is determined by the energy minimum. The quasiparticle condensation deeply influences the distribution of real particles. Apart from the original BEC in the $\k=\bold0$ one-particle state, two more condensates appear, a dominant one in the plane wave state $\k_\v$ and another one of a smaller density at $-\k_\v$. The coexistence of these condensates leads to density modulations characterized by the wave vectors $\k_\v$ and $ 2\k_\v$. In the present model, the density of the condensate at $\k_\v$ attains the full density at a finite velocity $v_1$; necessarily, the condensates at $\k=\bold0$ and $\k=-\k_\v$ and the two density modulations together with them vanish here, if not already at smaller velocities. At $v_1$ our model exhibits a bifurcation of $\k_\v$, with one solution increasing and the other one decaying as $v$ tends to infinity. The bifurcation is due to the fact that $\k_\v$ is determined by density saturation, when $v>v_1$. A general ground state is then a superposition of the two pure condensates, corresponding to the two solutions for $\k_\v$. There is no density modulation in these states.

\section*{Acknowledgement}

We thank Gergely Szirmai for numerical assistance. This work was supported by the Hungarian Science Foundation through OTKA Grant No. 77629.

\end{document}